# A Variable-Spot-Size and Multi-Frequency Square-Pulsed Source (SPS) Approach for Comprehensive Characterization of Anisotropic Thermal Transport Properties in Multilayered Thin Films

**Kexin Zhang[1], Tao Chen[1], Jinlong Ma[1,*], Puqing Jiang[1,*]**

[1]School of Energy and Power Engineering, Huazhong University of Science and Technology, Wuhan, Hubei 430074, China

## Abstract

Multilayered thin-film structures are frequently encountered in industrial applications, where accurate thermal property characterization is essential for performance optimization. These films, typically ranging from nanometers to micrometers in thickness, often exhibit anisotropic thermal conductivity and non-bulk heat capacity, which are challenging to measure. In this study, we introduce a variable-spot-size and multi-frequency square-pulsed source (SPS) method for the simultaneous determination of anisotropic thermal conductivities, heat capacities, and interfacial thermal conductance in multilayered systems. By leveraging a broad modulation frequency range (1 Hz to 10 MHz) and tunable laser spot sizes, the SPS method enhances sensitivity to different thermal parameters across layers. We validate this approach on a silicon-on-insulator (SOI) sample comprising a 1.59 μm Si layer, 1.03 μm SiO$_2$ layer, and a silicon substrate with a 122 nm aluminum (Al) transducer. The SPS method successfully extracts seven key thermal parameters, including the in-plane and cross-plane thermal conductivities and heat capacity of the Si film, the thermal conductivity and heat capacity of the SiO$_2$ layer, the thermal conductivity of the substrate, and the interfacial thermal conductance between Al and Si. Temperature-dependent measurements from 80 to 500 K showed excellent agreement with literature values and first-principles predictions, confirming the method's accuracy and reliability. These results demonstrate the SPS method as a powerful tool for comprehensive thermal characterization of complex multilayered structures, with implications for both fundamental research and practical applications.

*Corresponding Author: majinlong@hust.edu.cn; jpq2021@hust.edu.cn

# 1. Introduction

Thin-film structures, particularly those in the nanometer to micrometer range, play a critical role in modern industrial applications, ranging from microelectronics [1, 2] and optoelectronics [3] to thermal management in advanced devices [4]. These films often exhibit complex thermal behavior, such as anisotropic thermal conductivity [5] and deviations in volumetric heat capacity from bulk values [6]. Accurate measurement of these thermal properties is essential for device performance and reliability. However, conventional techniques such as calorimetry and steady-state methods are often inadequate for thin films, especially in multilayered systems where heat transfer is strongly influenced by interfacial effects and anisotropy.

Among existing approaches, pump-probe techniques such as time-domain thermoreflectance (TDTR) [7, 8] and frequency-domain thermoreflectance (FDTR) [9] have become popular for characterizing thermal properties at the micro- and nanoscale. These methods use a modulated pump laser to heat the sample and a probe laser to monitor temperature-induced reflectance changes. While powerful, their limited frequency range and sensitivity trade-offs often prevent comprehensive extraction of thermal conductivities (both in-plane and cross-plane), heat capacities, and interfacial thermal transport properties in layered film systems within a single framework.

To overcome these challenges, we introduce the square-pulsed source (SPS) method—a versatile pump-probe technique capable of comprehensive thermal characterization [10-12]. By using a square-wave modulated pump beam and analyzing time-resolved reflectance signals over a broad frequency range (1 Hz to 10 MHz) and varied laser spot sizes, the SPS method enhances sensitivity to both in-plane and cross-plane transport across multiple layers.

In this work, we apply the SPS method to a silicon-on-insulator (SOI) sample to demonstrate its effectiveness. The method enables the simultaneous extraction of seven key thermal parameters, including the in-plane and cross-plane thermal conductivities and heat capacity of the Si film, thermal conductivity and heat capacity of the $SiO_2$ layer, thermal conductivity of the substrate, and interfacial thermal conductance between Al and Si. Results over 80-500 K show excellent agreement with literature and theoretical predictions, validating the SPS method as an accurate and versatile tool for multilayer thin-film thermal characterization.

# 2. Methodologies

## 2.1. Basic principle of the SPS technique

The SPS technique, illustrated schematically in Fig. 1(a), employs a pump laser modulated with a 50% duty cycle square wave to generate periodic heating on the sample surface. A probe laser, focused on the same spot, detects the resulting temperature variations through the thermoreflectance effect. The reflected probe laser beam is captured by a photodetector, which converts the optical signals into corresponding electrical signals. These signals are processed by a periodic waveform analyzer, an integrated advanced function of the Zurich UHF lock-in amplifier, to produce time-resolved voltage signals over a complete square-wave heating cycle with a high signal-to-noise ratio. The resulting voltage signals, directly proportional to the sample's temperature response, undergo normalization along both amplitude and time axes. Finally, a thermal model is applied to best fit the normalized data, enabling accurate determination of the specimen's thermal properties [10, 13].

In our system, the pump and probe lasers operate at 458 nm and 785 nm, respectively. To conduct the measurements, the sample is typically coated with a ~100 nm thick aluminum (Al) film as a transducer layer. With the Al film thickness being significantly greater than its optical skin depth (~6.2 nm at 458 nm), all incident optical energy is absorbed near the surface. Therefore, heat generation is confined within the Al film, and thermal transport into the underlying layers occurs solely through diffusion. This assumption underpins the surface heating boundary condition used in our thermal model fitting.

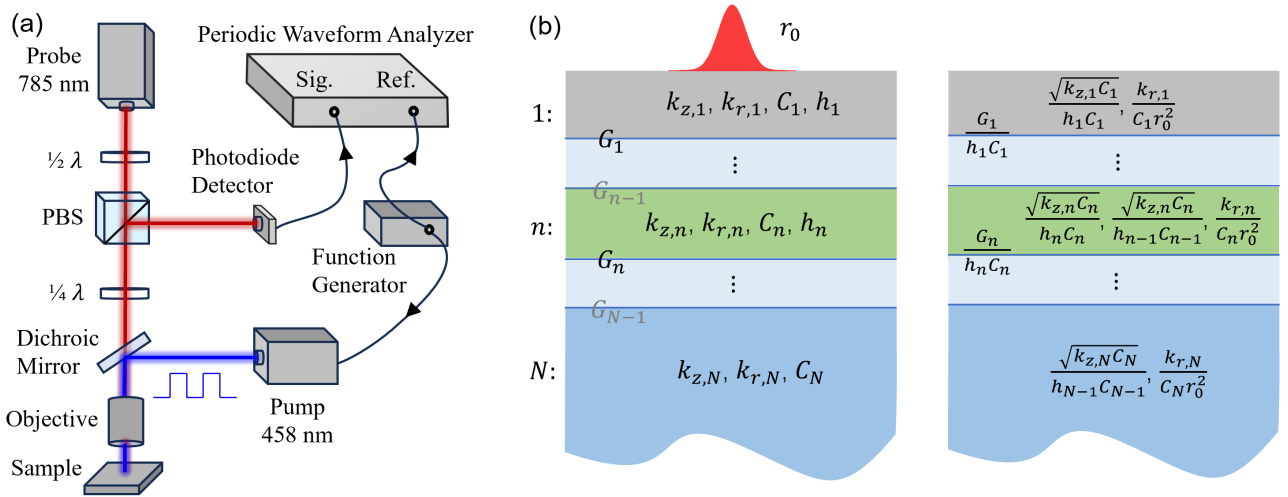

**Figure 1.** (a) Schematic diagram of the experimental setup for the SPS method. (b) Schematic illustrating individual and combined thermal parameters in an $N$-layered thin-film structure. The figure highlights how different parameters contribute to the system's thermal response. These parameters form the basis for sensitivity analysis and signal fitting in SPS measurements.

*2.2. Principle of measuring multiple parameters in multilayered systems*

The thermal response of an $N$-layered system is influenced by several parameters, including the in-plane thermal conductivity ($k_r$), cross-plane thermal conductivity ($k_z$), volumetric heat capacity ($C$), and thickness ($h$) of each layer, as well as the thermal conductance ($G$) at each interface. Additional factors include the laser spot size ($r_0$) and the modulation frequency ($f_0$), resulting in a total of $5N$ parameters. Among these, $f_0$ is a set parameter, while all others are subject to uncertainties. These parameters are summarized in Fig. 1(b).

The sensitivity of the normalized amplitude signal $A_{\text{norm}}$ to any parameter $\xi$ is quantified using the sensitivity coefficient, defined as:

$$S_\xi = \frac{\partial \ln A_{\text{norm}}}{\partial \ln \xi} = \frac{\xi}{A_{\text{norm}}} \frac{\partial A_{\text{norm}}}{\partial \xi} \tag{1}$$

This indicates that a 1% change in $\xi$ results in an $S_\xi$% change in the signal $A_{\text{norm}}$.

Chen and Jiang [14, 15] have demonstrated that the sensitivity coefficients for different parameters of the thermal system satisfy the following $N + 2$ relationships:

$$S_{C_n} = S_{h_n} + S_{k_{z,n}} - S_{k_{r,n}} \quad \text{for } n = 1,2, \ldots, N \tag{2}$$

$$\sum_{n=1}^{N-1} \left( S_{G_n} + S_{h_n} \right) = -2 \sum_{n=1}^{N} S_{k_{z,n}} \tag{3}$$

$$S_{r_0} = -2 \sum_{n=1}^{N} S_{k_{r,n}} \tag{4}$$

Building on these relationships, Chen and Jiang [14, 15] propose that the thermal response is governed by specific key combined parameters. For the $n$th layer, these parameters include $\frac{\sqrt{k_{z,n}C_n}}{h_n C_n}$, $\frac{\sqrt{k_{z,n}C_n}}{h_{n-1}C_{n-1}}$, and $\frac{k_{r,n}}{C_n r_0^2}$. Additionally, for the interface beneath the $n$th layer, the governing parameter is $\frac{G_n}{h_n C_n}$. These combined parameters, totaling $4N - 3$, are summarized in Fig. 1(b).

The sensitivity coefficients of $A_{\text{norm}}$ to these combined parameters can be derived from the sensitivity coefficients of the individual parameters as follows:

$$S_{\frac{\sqrt{k_{z,1}C_1}}{h_1 C_1}} = 2S_{k_{z1}} \tag{5}$$

$$S_{\frac{\sqrt{k_{zn}C_n}}{h_n C_n}} = 2S_{k_{z,n}} + \sum_{l=1}^{l=n-1} \left( 2S_{k_{z,l}} + S_{h_l} + S_{G_l} \right); \quad \text{for } n = 2, \ldots, N-1 \tag{6}$$

$$S_{\frac{\sqrt{k_{zn}C_n}}{h_{n-1}C_{n-1}}} = - \sum_{l=1}^{l=n-1} \left( 2S_{k_{zl}} + S_{h_l} + S_{G_l} \right); \quad \text{for } n = 2, \ldots, N \tag{7}$$

$$S_{\frac{G_n}{h_n C_n}} = S_{G_n}; \quad \text{for } n = 1, \dots, N-1 \tag{8}$$

$$S_{\frac{k_{rn}}{C_n r_0^2}} = S_{k_{rn}}; \quad \text{for } n = 1, \dots, N \tag{9}$$

These combined parameters not only reveal the relationships among various parameters but also define the maximum number of parameters that can be extracted through fitting experimental signals. In some instances, certain combined parameters may exhibit partial coupling. To quantitatively evaluate the number of measurable parameters, singular value decomposition (SVD) can be performed on the sensitivity matrix: $\boldsymbol{S} = \left[ S_{\xi_1} \ \dots \ S_{\xi_j} \ \dots \ S_{\xi_q} \right]_{p \times q}$, where $p$ is the number of data points and $q$ is the number of combined parameters being analyzed. Each column, $S_{\xi_j} = (S_{\xi_j, t_1}, \dots, S_{\xi_j, t_p})^T$, represents the sensitivity of the signal $A_{\text{norm}}$ to the combined parameter $\xi_j$ at different time points. The number of measurable parameters is determined by the number of singular values of the sensitivity matrix $\boldsymbol{S}$ that are larger than 0.1. These theories will be demonstrated through the measurement example of an SOI sample in Section 3.

# 3. Results and discussion

## 3.1. Sample preparation and structure characterization

The sample in this study is a commercial SOI wafer fabricated using the Smart-Cut[TM] process, with a nominal device layer thickness of 1.5 $\mu$m and an oxide layer thickness of 1 $\mu$m. For thermal measurements, a 100 nm Al transducer layer was deposited via electron beam evaporation. As shown in the cross-sectional SEM image in Figure 2(a), the actual measured thicknesses of the Al, Si, and SiO$_2$ layers are $122 \pm 6$ nm, $1.59 \pm 0.05$ $\mu$m, and $1.03 \pm 0.03$ $\mu$m, respectively. The 122 nm Al thickness ensures complete absorption of the pump and probe beams within the Al layer, consistent with the surface heating assumption used in our thermal model (see Section 2.1). The SEM image also reveals stratified sublayers within the SiO$_2$ layer. Further elemental analysis using Energy Dispersive X-ray Spectroscopy (EDX), shown in Fig. 2(b), suggests that this stratification is likely structural or morphological rather than compositional.

A multilayered thermal model based on the sample structure is developed, as shown in Fig. 2(c). This model involves 19 intrinsic parameters, including $k_r$, $k_z$, $C$, and $h$ of the metal transducer layer (Al), device layer (Si film), and oxide layer (SiO$_2$). It also accounts for $k_r$, $k_z$, and $C$ of the bulk Si substrate. Three interfacial thermal conductances at the Al/Si film interface ($G_1$), the Si film/SiO$_2$ interface ($G_2$), and the SiO$_2$/bulk Si interface ($G_3$) are involved, in addition to the laser spot radius ($r_0$). As discussed in Section 2, these individual parameters combine to form 13 key parameters that primarily govern the heat diffusion process in the SOI sample, as shown in Fig. 2(c). For clarity, these combined parameters are labeled with double-digit circled numbers, where the first digit represents the layer number and the second indicates the parameter's sequence within that layer.

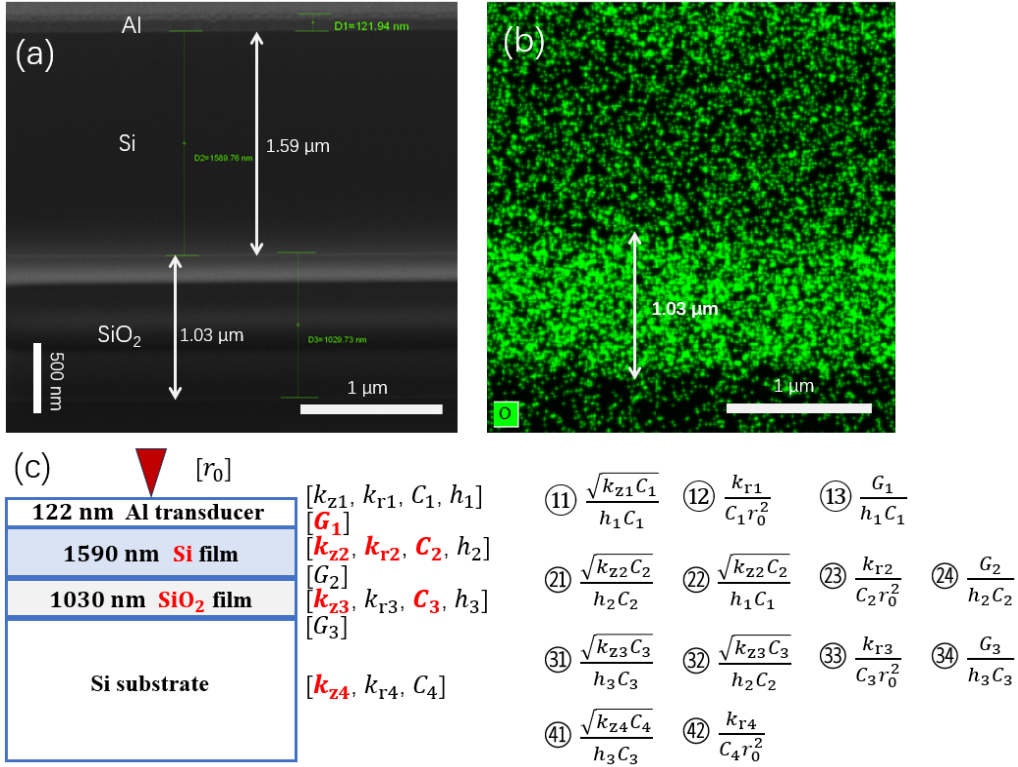

**Figure 2.** (a) Cross-sectional SEM image of the SOI sample, showing distinct layers of the Al transducer, Si device layer, and buried SiO₂. (b) EDX mapping of the SiO₂ layer, confirming uniform elemental distribution and indicating that the visible stratification arises from morphological, not compositional, variations. (c) Thermal model schematic used in SPS signal fitting, incorporating 19 intrinsic parameters and their grouping into 13 key combined parameters that govern heat diffusion. The diagram also defines the labeling system for these combined parameters, which are central to sensitivity and SVD analysis.

### 3.2. Measurement procedure

We employed a variable spot size and multi-frequency approach based on the SPS technique to probe multiple thermal parameters of the multilayered system. Specifically, high modulation frequencies (in the MHz range) were used to probe the thermal properties of the topmost layers, while low modulation frequencies (in the kHz range) probed the substrate. A large laser spot size reduced sensitivity to in-plane thermal transport, whereas a small spot size enhanced it. To illustrate the measurement process, we present an example of room-temperature measurements on this multilayered SOI system.

### 3.2.1 Qualitative analysis based on sensitivity curves

Figure 3(1a-1f) shows the SPS signals measured at a spot size of 14.4 $\mu$m and a modulation frequency of 2 MHz, along with the corresponding sensitivity coefficients. Figures 3(1a) and 3(1b) display the heating and cooling phase signals, where the symbols represent experimental data and the curves represent best-fit results from the thermal model. Figures 3(1c) and 3(1d) plot the sensitivity coefficients of the signals in Figures 3(1a) and 3(1b) for all 19 individual parameters. For clarity, only parameters with sensitivity magnitudes greater than 0.05 are labeled.

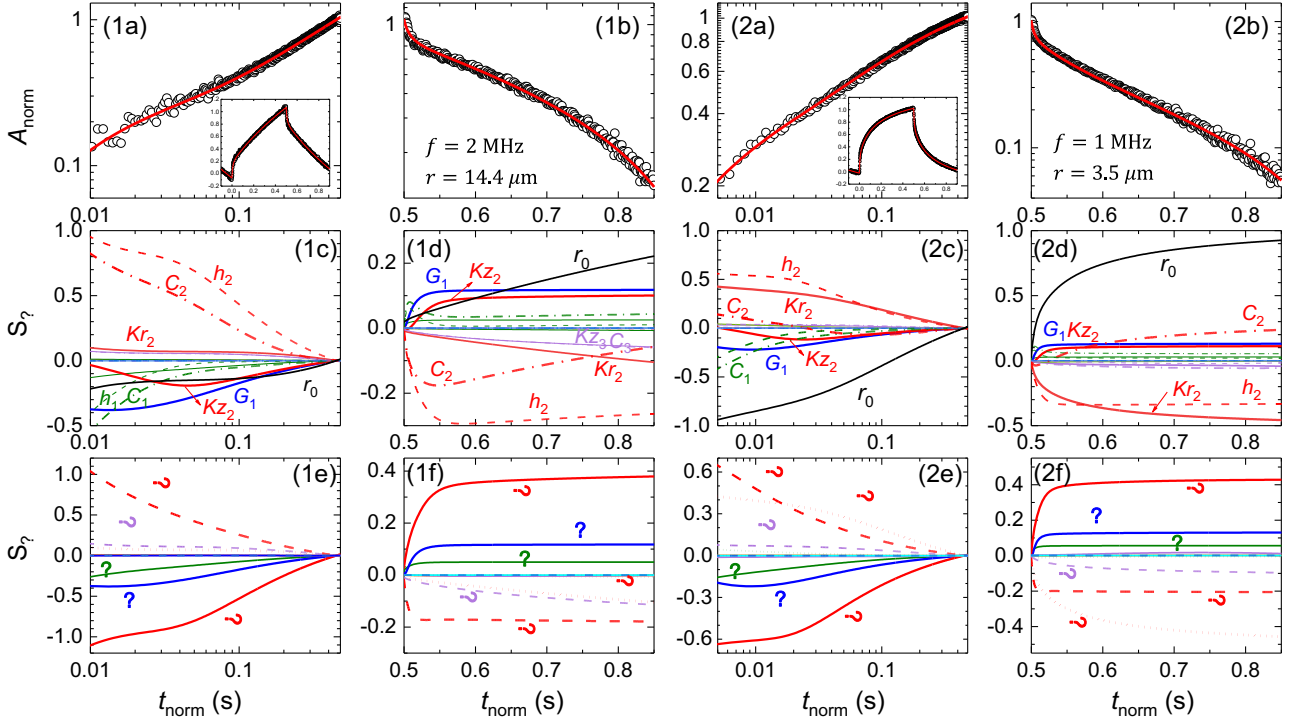

**Figure 3.** SPS signals and sensitivity analyses at MHz-range frequencies used to extract thermal properties of the Si device layer.

(1a-1f) show results for a 2 MHz modulation frequency and a 14.4 $\mu$m laser spot size, primarily probing $G_1$, $k_{z2}C_2$, and $C_2$ of the Si layer. (2a-2f) show results at 1 MHz with a smaller 3.5 $\mu$m spot size, enhancing sensitivity to $k_{r2}/C_2$. The associated sensitivity coefficients indicate which parameters most affect the signal, guiding which properties can be reliably extracted. The combined parameters include ⑪ $\frac{\sqrt{k_{z1}C_1}}{h_1C_1}$, ⑬ $\frac{G_1}{h_1C_1}$, ㉑ $\frac{\sqrt{k_{z2}C_2}}{h_2C_2}$, ㉒ $\frac{\sqrt{k_{z2}C_2}}{h_1C_1}$, ㉓ $\frac{k_{r2}}{C_2r_0^2}$, and ㉜ $\frac{\sqrt{k_{z3}C_3}}{h_2C_2}$.

The sensitivity analysis indicates that this set of signals is most sensitive to the properties of the device layer, particularly the cross-plane thermal conductivity $k_{z2}$, volumetric heat capacity $C_2$, and thickness $h_2$. Some parameters exhibit partial correlations, such as $S_{r_0} = -2S_{k_{r2}}$, and $S_{C_2} = S_{k_{z2}} - S_{k_{r2}} + S_{h_2}$.

Figures 3(1e) and 3(1f) present the sensitivity coefficients to combined parameters, revealing that the signals are primarily sensitive to the following combined terms: ⑬$\frac{G_1}{h_1 C_1}$, ㉑$\frac{\sqrt{k_{z2} C_2}}{h_2 C_2}$, and ㉒$\frac{\sqrt{k_{z2} C_2}}{h_1 C_1}$. A closer examination shows that these combined parameters are not coupled, suggesting that they can be independently determined by fitting this set of signals. Once these combined parameters are extracted, we can determine $G_1$ from $\frac{G_1}{h_1 C_1}$, $k_{z2} C_2$ from $\frac{\sqrt{k_{z2} C_2}}{h_1 C_1}$, and $C_2$ from $\frac{\sqrt{k_{z2} C_2}}{h_2 C_2}$, provided that $h_1$, $h_2$ and $C_1$ are known.

To enhance sensitivity to the in-plane thermal conductivity $k_{r_2}$ of the device layer, we repeated the measurements at a similar modulation frequency (1 MHz) but with a significantly smaller spot size of 3.5 $\mu m$. The corresponding signals and sensitivity coefficients are shown in Figure 3(2a-2f). The sensitivity plots in Figures 3(2e) and 3(2f) confirm that this set of signals remains sensitive to ⑬$\frac{G_1}{h_1 C_1}$, ㉑$\frac{\sqrt{k_{z2} C_2}}{h_2 C_2}$ and ㉒$\frac{\sqrt{k_{z2} C_2}}{h_1 C_1}$, but with a markedly increased sensitivity to ㉓$\frac{k_{r2}}{C_2 r_0^2}$ compared to the earlier measurement. Therefore, knowing $r_0$, $h_1$, $h_2$, and $C_1$, we can simultaneously fit both sets of signals to accurately determine $G_1$, $k_{z2}$, $C_2$, and $k_{r2}$.

To probe deeper layers, we employed a large laser spot size of 14.4 $\mu m$ and lowered the modulation frequency to 100 kHz and 10 kHz. The measured signals and corresponding sensitivity coefficients are shown in Figure 4. Lower modulation frequencies increase thermal penetration depth, making the signals more sensitive to underlying layers.

At 100 kHz, heat penetrates through the SiO$_2$ layer, and the signals exhibit strong sensitivity to the following combined parameters: ㉑$\frac{\sqrt{k_{z2} C_2}}{h_2 C_2}$, ㉒$\frac{\sqrt{k_{z2} C_2}}{h_1 C_1}$, ㉓$\frac{k_{r2}}{C_2 r_0^2}$, ㉛$\frac{\sqrt{k_{z3} C_3}}{h_3 C_3}$, and ㉜$\frac{\sqrt{k_{z3} C_3}}{h_2 C_2}$, as shown in Figures 4(1e) and 4(1f). Since the device layer parameters have already been determined from the higher-frequency measurements, the sensitivity curves indicate that parameters ㉛ and ㉜ –associated with the oxide layer–are not coupled. Therefore, with the SiO$_2$ layer thickness $h_3$ known, the cross-plane thermal conductivity $k_{z3}$ and volumetric heat capacity $C_3$ of the oxide layer can be extracted from these signals.

Further lowering the modulation frequency to 10 kHz increases sensitivity to the substrate properties. As shown in Figures 4(2e) and 4(2f), the signals (particularly in the cooling phase) show strong sensitivity to: �француз㊶ $\frac{\sqrt{k_{z4} C_4}}{h_3 C_3}$ and ㊷$\frac{k_{r4}}{C_4 r_0^2}$, parameters related to the substrate. However, Figures 4(2c) and 4(2d) show that the volumetric heat capacity $C_4$ of the substrate has very low sensitivity compared to other intrinsic parameters. This implies that treating $C_4$ as an unknown would introduce significant fitting uncertainty. Therefore, $C_4$ should be treated as a known input. With $h_3$ and $C_3$ known, $k_{z4}$ of the Si substrate can be determined by fitting this final set of signals.

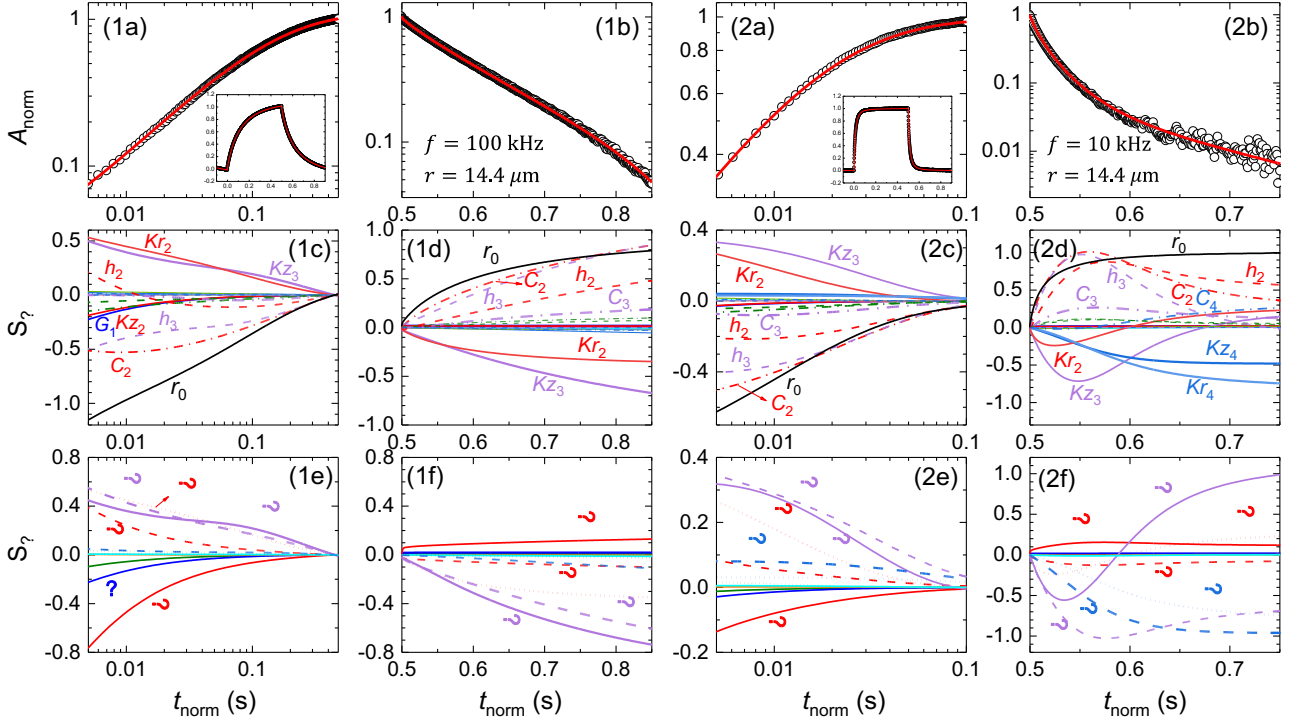

**Figure 4.** SPS signals and sensitivity analyses at lower kHz-range frequencies used to probe deeper layers, including the buried oxide layer and the Si substrate.

(1a-1f) show results at 100 kHz, capturing the thermal response of the oxide layer. (2a-2f) show results at 10 kHz, increasing sensitivity to the Si substrate. The evolving sensitivity coefficients with decreasing modulation frequency illustrate how thermal penetration depth shifts with frequency, enabling selective probing of different layers. The combined parameters include ⑬ $\frac{G_1}{h_1 C_1}$, ㉑ $\frac{\sqrt{k_{z2} C_2}}{h_2 C_2}$, ㉒ $\frac{\sqrt{k_{z2} C_2}}{h_1 C_1}$, ㉓ $\frac{k_{r2}}{C_2 r_0^2}$, ㉛ $\frac{\sqrt{k_{z3} C_3}}{h_3 C_3}$, ㉜ $\frac{\sqrt{k_{z3} C_3}}{h_2 C_2}$, ㊶ $\frac{\sqrt{k_{z4} C_4}}{h_3 C_3}$, and ㊷ $\frac{k_{r4}}{C_4 r_0^2}$.

### 3.2.2 Quantitative evaluation based on SVD analysis

We further assessed the measurability of the thermal parameters using SVD analysis. To apply this method, the sensitivities of the combined parameters under investigation were arranged into the matrix $\mathbf{S}$ as follows:

$$\mathbf{S} = \begin{pmatrix} S_{(1,\xi_1)} & \cdots & S_{(1,\xi_8)} \\ \vdots & \ddots & \vdots \\ S_{(4,\xi_1)} & \cdots & S_{(4,\xi_8)} \end{pmatrix}_{4p \times 8} \tag{10}$$

Here, $\mathbf{S}_{(i,\xi_j)}$ is a column vector of size $p \times 1$ (where $p$ is the number of data points in each signal set), containing the sensitivity coefficients of the $i$th set of signals to the $j$th combined parameter. The eight combined parameters analyzed were: $\frac{G_1}{h_1 C_1}$, $\frac{\sqrt{k_{z2} C_2}}{h_2 C_2}$, $\frac{\sqrt{k_{z2} C_2}}{h_1 C_1}$, $\frac{k_{r2}}{C_2 r_0^2}$, $\frac{\sqrt{k_{z3} C_3}}{h_3 C_3}$, $\frac{\sqrt{k_{z3} C_3}}{h_2 C_2}$, $\frac{\sqrt{k_{z4} C_4}}{h_3 C_3}$, and $\frac{k_{r4}}{C_4 r_0^2}$. The other five combined parameters, whose sensitivities remained below 0.05 across all four signal sets, were excluded from the analysis.

Upon decomposition, the diagonal matrix was obtained as:

$$\Sigma = \text{diag}(55.94, 28.58, 18.46, 7.41, 4.93, 1.49, 1.05, 0.63)$$

Since all singular values exceeded 0.1, it confirms that these eight combined parameters are mathematically uncoupled. Therefore, in theory, eight thermal parameters—$G_1$, $k_{z2}$, $C_2$, $k_{r2}$, $k_{z3}$, $C_3$, $k_4$, and $C_4$—can be independently determined by fitting the four sets of signals, assuming prior knowledge of $h_1$, $C_1$, $h_2$, $h_3$, and $r_0$, as well as the substrate isotropy condition $k_4 = k_{r4} = k_{z4}$.

However, when attempting to fit all eight parameters simultaneously, we observed unacceptably high uncertainty in $C_4$. This is attributed to the relatively weak sensitivity of $C_4$, as discussed in Section 3.2.1. Therefore, in practice, we treat $C_4$ as a known input parameter.

This example demonstrates that SVD analysis is an effective tool for identifying the number of independently measurable parameters and assessing parameter coupling. However, while SVD establishes the mathematical independence of parameters, their practical measurability ultimately depends on the relative magnitude of their sensitivities.

### 3.2.3 Best fitting results

Building on the qualitative and quantitative analyses above, we conclude that seven thermal parameters of the SOI thermal system—$G_1$, $k_{z2}$, $C_2$, $k_{r2}$, $k_{z3}$, $C_3$, and $k_4$—can be simultaneously determined by fitting the four sets of signals shown in Figures 3 and 4. Meanwhile, six parameters of the system—$h_1$, $C_1$, $h_2$, $h_3$, $C_4$, and $r_0$—must be accurately known as input values. The remaining parameters ($k_{r1}$, $k_{z1}$, $G_2$, $k_{r3}$, and $G_3$) are also treated as inputs, but their precise values have negligible impact on the fitting results.

Among the input parameters, the layer thicknesses were measured from high-resolution SEM cross-sectional images. The temperature-dependent heat capacities $C_1$ and $C_4$ were obtained from the TPRC database, with room-temperature values of 2.44 J/(cm$^3$ · K) for Al and 1.66 J/(cm$^3$ · K) for bulk Si. The laser spot sizes for each measurement set were determined using the spatial-domain thermoreflectance (SDTR) method [16], with an estimated uncertainty of 2%. The in-plane thermal conductivity of the Al film ($k_{r1}$) at room temperature was calculated from its electrical resistivity (measured via the van der Paur method) using the Wiedemann-Franz law.

The temperature dependence of $k_{r1}$ was inferred from bulk single-crystal Al data [17], assuming a constant offset between bulk and thin-film electrical resistivities. Isotropic thermal conductivities were assumed for the Al film, SiO$_2$ film, and Si substrate, with $k_{r1} = k_{z1}$, $k_{r3} = k_{z3}$, and $k_{r4} = k_{z4}$. The interfacial thermal conductances of the Si/SiO$_2$ and SiO$_2$/Si interfaces were assumed as $G_2 = G_3 = 200 \pm 100$ MW/(m$^2 \cdot$ K); sensitivity analysis confirmed that variations in these values had negligible effect on the fitted results.

To extract the seven fitted parameters, we employed a hybrid optimization algorithm combining particle swarm optimization (PSO) with the quasi-Newton method to fit the four signal sets in Figures 3 and 4. This hybrid approach achieved not only improved accuracy but also significantly higher computational efficiency than the commonly used MATLAB function "lsqnonlin". Briefly, PSO—a global optimization scheme inspired by swarm intelligence—explores the parameter space to identify global optima. While highly robust and effective at avoiding local minima in high-dimensional spaces, PSO's convergence rate declines after ~10 iterations. To address this limitation, we transitioned to the quasi-Newton method, a gradient-based local optimizer, which rapidly refines the solution to high precision. By integrating the global exploration strength of PSO with the fast local convergence of quasi-Newton, we achieve efficient and accurate parameter estimation.

**Table 1** Room-temperature thermal properties of the SOI sample measured using the SPS method.

| Layer/Interface | Parameter | Value $\pm$ Uncertainty | Unit | Error (%) |
|---|---|---|---|---|
| **Al/Si Interface** | $G_1$ | $160 \pm 10.9$ | MW/(m$^2 \cdot$ K) | 6.8 |
| **Si Film** | $k_{z2}$ (cross-plane) | $100 \pm 4.7$ | W/(m$\cdot$K) | 4.6 |
| | $k_{r2}$ (in-plane) | $116 \pm 7.3$ | W/(m$\cdot$K) | 6.3 |
| | $C_2$ | $1.66 \pm 0.06$ | J/(cm$^3 \cdot$ K) | 4.0 |
| **SiO$_2$ Film** | $k_{z3}$ | $1.38 \pm 0.07$ | W/(m$\cdot$K) | 4.9 |
| | $C_3$ | $1.65 \pm 0.09$ | J/(cm$^3 \cdot$ K) | 5.5 |
| **Si Substrate** | $k_4$ (isotropic) | $148 \pm 6$ | W/(m$\cdot$K) | 4.0 |

The best-fit results are as follows: the Al/Si interfacial thermal conductance is $G_1 = 160 \pm 10.9$ MW/(m$^2 \cdot$ K); the cross-plane and in-plane thermal conductivity of the Si film are $k_{z2} = 100 \pm 4.7$ W/(m$\cdot$K) and $k_{r2} = 116 \pm 7.3$ W/(m$\cdot$K), respectively; the volumetric heat capacity of the Si layer is $C_2 = 1.66 \pm 0.06$ J/(cm$^3 \cdot$ K); the thermal conductivity and volumetric heat capacity of the SiO$_2$ layer are $k_{z3} = 1.38 \pm 0.07$ W/(m$\cdot$K) and $C_3 = 1.65 \pm 0.09$ J/(cm$^3 \cdot$ K); and the thermal conductivity of the Si substrate is $k_{z4} = 148 \pm 6 \, \mathrm{W/(m \cdot K)}$. Uncertainties were estimated using a full error propagation method that accounts for both the uncertainties of input parameters and signal noise. This uncertainty analysis approach, originally developed by Yang et al. [18], has been widely adopted in studies employing TDTR, FDTR, and other thermoreflectance techniques [10, 11, 16, 19-22].

### 3.3. Temperature-dependent results and discussion

The measurements were repeated across temperatures from 80 to 500 K, with results summarized in Figure 5 as a function of temperature. Solid symbols represent the current measurements, solid curves denote recommended values from the TPRC database (for $k_{\mathrm{Si,bulk}}$, $k_{\mathrm{SiO_2}}$, $C_{\mathrm{Si}}$, and $C_{\mathrm{SiO_2}}$), and dashed curves correspond to theoretical predictions (for $k_{r,\mathrm{Si\,film}}$, $k_{z,\mathrm{Si\,film}}$, and $G_{\mathrm{Al/Si}}$). The observed temperature dependence of the thermal parameters offers important insight into the underlying phonon transport mechanisms and the structural characteristics of the multilayered SOI system.

For parameters with well-established literature values ($k_{\mathrm{Si,bulk}}$, $k_{\mathrm{SiO_2}}$, $C_{\mathrm{Si}}$, and $C_{\mathrm{SiO_2}}$), our measurements show excellent agreement with TPRC recommendations [23, 24]. Notably, as the temperature decreases toward 100 K, the measured $k_{\mathrm{Si,bulk}}$ increases to 890 W/(m · K), aligning with accepted literature despite an uncertainty of up to 60%. This large uncertainty primarily stems from the reduced signal-to-noise ratio (SNR) under cryogenic conditions, where the substrate exhibits significantly higher thermal conductivity.

To ensure sufficient sensitivity to $k_{\mathrm{Si,bulk}}$ at low temperatures, we employ a large laser spot size ($r_0 > 30 \, \mu\mathrm{m}$) and a low modulation frequency ($f_0 < 20 \, \mathrm{kHz}$). However, this combination reduces the magnitude of the surface temperature rise, necessitating higher pump laser power to maintain an adequate SNR. Our current system is limited to a maximum pump power of 150 mW, which is sufficient for most measurements but constrains the achievable SNR under these specific conditions.

As with other thermoreflectance techniques, the SPS method requires that the steady-state surface temperature rise induced by laser heating remains below 10 K or 10% of the sample's absolute temperature, whichever is smaller [7], to ensure linearity and avoid sample damage. The steady-state surface temperature rise for a multilayered system can be estimated using [25]:

$$\Delta T = \frac{P}{\pi r_0^2} \sum_{i=1}^{N-1} \left( \frac{h_i}{k_{z,i}} + \frac{1}{G_i} \right) + \frac{P}{2\sqrt{\pi r_0^2 k_{r,N} k_{z,N}}} \tag{11}$$

where $P = \alpha_1 P_1 + \alpha_2 P_2$ is the total absorbed laser power from the pump and probe beams, with absorptivities $\alpha_1$, $\alpha_2$, and powers $P_1$, $P_2$. This equation shows that inadequate absorbed power may lead to a sub-optimal temperature rise, while excessive laser power risks non-linear effects or damage.

The absorptivity of the transducer is a key factor in this balance. In our case, the Al transducer absorbs only ~10% of the incident light at 458 nm. Replacing it with a higher-absorptivity material such as titanium nitride (TiN), which absorbs up to ~80% at similar wavelengths [26], could enhance thermal excitation efficiency without increasing laser power. This improvement would help reduce uncertainty in low-temperature measurements and expand the SPS method's applicability to high-conductivity or thermally constrained samples.

In contrast, measurements of the Si and SiO₂ films are less affected by these constraints at low temperatures, as a smaller spot size ($r_0 \approx 5 \mu m$) remains sufficient for sensitivity. The thermal conductivity of the SiO₂ layer $k_{SiO_2}$ remains low across all temperatures due to its amorphous structure, which limits phonon mean free paths.

Both the measured $C_{Si}$ and $C_{SiO_2}$ values closely follow TPRC recommendations [23] and increase with temperature, as expected from the Debye model, which predicts a cubic temperature dependence at low temperatures and saturation at high temperatures. Notably, $C_{Si}$ and $C_{SiO_2}$ are comparable across the measured temperature range. Between 100 K and 300 K, $C_{Si}$ is slightly higher than $C_{SiO_2}$, whereas above 300 K, this trend reverses, with $C_{Si}$ becoming slightly lower—a subtle behavior that is consistent with TPRC reference data. The heat capacity of the Si film closely follows the bulk trend, indicating minimal deviation from bulk phonon density of states despite its sub-micrometer thickness.

For the other three parameters ($k_{r,Si\,film}$, $k_{z,Si\,film}$, and $G_{Al/Si}$) lacking widely established literature values, our measurements also align well with theoretical predictions. Here, predictions of $k_{r,Si\,film}$ and $k_{z,Si\,film}$ for the 1.6 μm-thick Si film were computed using the first-principles method (see Appendix A1 for calculation details), and $G_{Al/Si}$ was obtained from reference [27], where it was calculated based on the diffuse mismatch model (DMM). Both $k_{r,Si\,film}$ and $k_{z,Si\,film}$ are lower than the bulk thermal conductivity and increase as temperature decreases. Their difference is small above room temperature but becomes more pronounced at lower temperatures, all matching theoretical expectations.

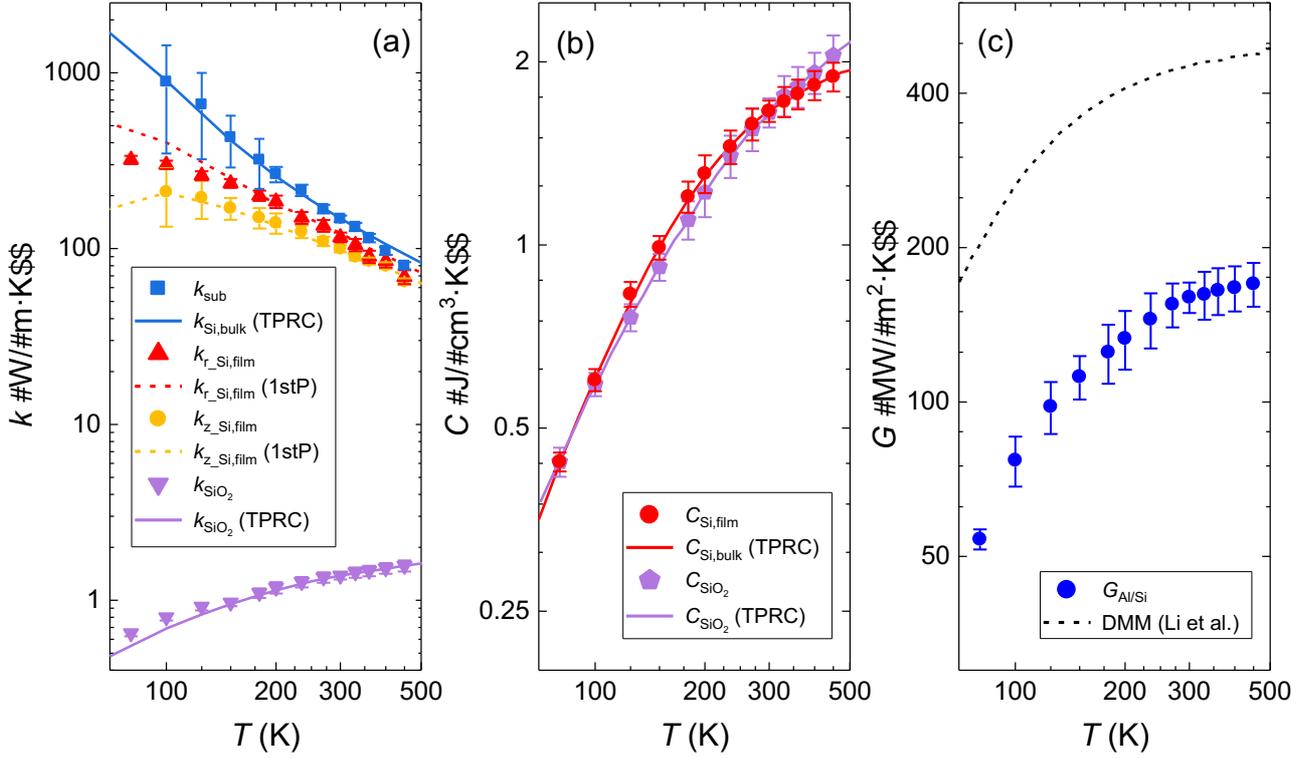

**Figure 5.** Temperature dependence of the seven thermal parameters extracted from the SOI sample using the SPS method: (a) In-plane and cross-plane thermal conductivities of the Si film ($k_{r,\text{Si film}}$ and $k_{z,\text{Si film}}$), isotropic thermal conductivities of the SiO2 layer ($k_{\text{SiO}_2}$) and the Si substrate ($k_{\text{Si,bulk}}$); (b) Volumetric heat capacities of the Si ($C_{\text{Si}}$) and SiO2 ($C_{\text{SiO}_2}$) layers; and (c) Interfacial thermal conductance at the Al/Si interface ($G_{\text{Al/Si}}$). Solid symbols represent the SPS measurements at different temperatures, with vertical error bars indicating uncertainty based on full error propagation. Solid lines denote recommended values from the TPRC database [23, 24], and dashed lines represent first-principles or model-based predictions [27].

It is important to note that as the temperature decreases to 100 K, the uncertainty in $k_{z,\text{Si film}}$ from our measurements increases substantially, reaching 36%. This large uncertainty primarily results from the reduced sensitivity of the measurement to $k_{z,\text{Si film}}$, which is dictated by the sample's thermal structure. At 100 K, the combination of reduced $k_{\text{SiO}_2}$ and increased $k_{z,\text{Si film}}$ causes the cross-plane thermal resistance of the system to be dominated by the oxide layer. The thermal resistance of the Si film becomes negligible, rendering the surface temperature response insensitive to $k_{z,\text{Si film}}$, leading to significant uncertainty in its determination. In contrast, the measurement of $k_{r,\text{Si film}}$ remained unaffected, as the increased thermal resistance of the oxide layer enhances lateral heat spreading within the Si film, thereby maintaining high sensitivity to $k_{r,\text{Si film}}$.

The measured $G_{\text{Al/Si}}$ increases with temperature, aligning qualitatively with the DMM, though with consistently lower values. This discrepancy arises from the DMM's idealized assumption of an atomically sharp, oxide-free interface–an assumption that does not hold for our sample. Specifically, a 2-3 nm native $SiO_2$ layer exists at the Al/Si interface due to the absence of hydrofluoric acid (HF) cleaning before Al deposition on the SOI substrate. This interfacial oxide introduces additional phonon scattering mechanisms and thermal resistance not accounted for by the DMM, thereby limiting the model's quantitative accuracy in describing real-world systems.

In the literature, similar SOI structures were previously studied by Jiang et al. [28, 29] using TDTR to extract $k_{r,\text{Si film}}$ and $k_{z,\text{Si film}}$ of Si films with thicknesses ranging from 1-10 $\mu$m. Their approach employed a dual-frequency TDTR approach [30] with modulation frequencies of 10 MHz and 3 MHz to minimize sensitivity to the Al transducer layer, thereby improving the accuracy of $k_{z,\text{Si film}}$, with a reported uncertainty of approximately 10%. To measure $k_{r,\text{Si film}}$, a low modulation frequency of 0.5 MHz was used, with $k_{z,\text{Si film}}$ treated as a fixed input parameter. In contrast to our current study, which simultaneously determines seven thermal properties of the SOI sample, the TDTR measurements were limited to extracting only three: $k_{r,\text{Si film}}$, $k_{z,\text{Si film}}$, and $G_{\text{Al/Si}}$.

Figure 6 compares our measurements of $k_{r,\text{Si film}}$ and $k_{z,\text{Si film}}$ for the 1.6 $\mu$m-thick Si film with those reported by Jiang et al. [28, 29], alongside first-principles predictions at 300 K and 100 K. At both temperatures, our results show good agreement with both the TDTR data and theoretical values. Importantly, our measurements exhibit reduced uncertainties relative to TDTR, primarily because our method simultaneously determines a larger set of thermal parameters. As a result, uncertainties in intermediate parameters do not propagate into the final values of $k_{r,\text{Si film}}$ and $k_{z,\text{Si film}}$. Furthermore, while TDTR could only characterize relatively thick Si films ($> 4 \mu$m) at 100 K, our method remains effective for thinner films, such as the 1.6 $\mu$m-thick sample, due to the enhanced sensitivity provided by the variable-spot-size, multi-frequency SPS approach across a broad range of modulation frequencies and laser spot sizes. This comparison underscores several advantages of the SPS method over TDTR, including its ability to extract a larger set of thermal parameters simultaneously, reduced uncertainty in measured values, and superior applicability to thinner films and low-temperature conditions where TDTR becomes less effective. Overall, the excellent agreement between our measurements, prior experimental data, and first-principles predictions confirms the accuracy and robustness of the current method.

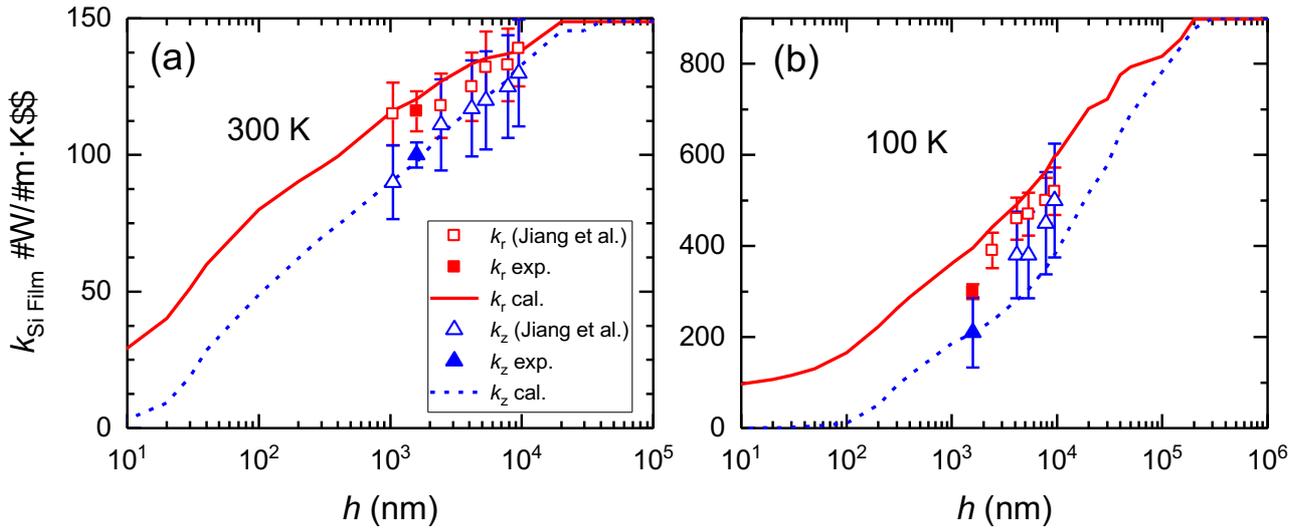

**Figure 6.** Comparison of the in-plane and cross-plane thermal conductivities of the 1.6 $\mu$m-thick Si film obtained using the SPS method with literature data and theoretical predictions at two temperatures: (a) 300 K and (b) 100 K. Solid symbols represent the SPS measurements from this work; open symbols correspond to TDTR results reported by Jiang et al. [28, 29]. Curves represent first-principles predictions. The SPS results show strong agreement with both experimental and theoretical references, with reduced uncertainty achieved through simultaneous multi-parameter fitting.

## 4. Conclusions

In summary, we have developed and validated a variable-spot-size, multi-frequency square-pulsed source (SPS) method for comprehensive thermal characterization of multilayered thin-film systems. This approach enables the simultaneous extraction of in-plane and cross-plane thermal conductivities, volumetric heat capacities, and interfacial thermal conductance within a single experimental framework. By combining a wide modulation frequency range with tunable laser spot sizes, the method achieves enhanced sensitivity across multiple layers and directions of heat transport.

Applied to a silicon-on-insulator (SOI) sample, the SPS method successfully determined seven key thermal parameters with high accuracy. These results exhibit strong consistency with literature data and theoretical predictions over a broad temperature range (80–500 K), confirming its reliability and broad applicability for analyzing multilayered thermal transport. These findings highlight the strength of the SPS technique in resolving anisotropic and interface-dominated heat transport in complex multilayer systems.

Although this study focused on an SOI sample, the SPS technique is broadly applicable to other multilayered systems, particularly those with anisotropy, buried interfaces, or non-bulk thermal properties. Its tunable measurement configuration offers flexibility across a range of material geometries and thermal regimes.

However, challenges may arise when dealing with strong lateral heterogeneity, non-planar geometries, or temperature-dependent optical properties, where accurate thermal modeling becomes more complex. Addressing these limitations will be an important direction for future refinement and application of the SPS method.

## Appendix

### A1. First-principles calculation of $k_r$ and $k_z$ of Si films

The lattice thermal conductivity tensor can be calculated from the phonon Boltzmann transport equation (BTE) as:

$$\boldsymbol{\kappa} = \frac{1}{\Omega N} \sum_{\mathbf{q}p} \hbar \omega_{\mathbf{q}p} \mathbf{v}_{\mathbf{q}p} \mathbf{F}_{\mathbf{q}p} \left( \frac{\partial n_{\mathbf{q}p}^0}{\partial T} \right) \tag{A1}$$

where $\mathbf{q}p$ denotes a phonon mode labeled by wave vector $\mathbf{q}$ and polarization $p$, $\Omega$ is the unit cell volume, and $N$ is the number of sampled points in the first Brillouin zone. $\omega_{\mathbf{q}p}$ and $\mathbf{v}_{\mathbf{q}p}$ are the phonon frequency and group velocity, respectively. $n_{\mathbf{q}p}^0$ is the Bose-Einstein distribution function, and $\mathbf{F}_{\mathbf{q}p}$ is the phonon mean free displacement [31], given by

$$\mathbf{F}_{\mathbf{q}p} = \tau_{\mathbf{q}p}^0 \left( \mathbf{v}_{\mathbf{q}p} + \Delta_{\mathbf{q}p} \right) \tag{A2}$$

Here, $\tau_{\mathbf{q}p}^0$ is the relaxation time accounting for both three-phonon and isotope scattering, and $\Delta_{\mathbf{q}p}$ is the correction term used in the iterative solution of the BTE. The second- and third-order interatomic force constants (IFCs) for Si are adopted from the *FourPhonon* package [31], and a 40×40×40 q-point mesh is employed. This setup yields a calculated bulk thermal conductivity of 148 W/(m · K) at room temperature.

For thin films, the relaxation time is further modified by a suppression function $S_{\mathbf{q}p}$ to account for additional boundary scattering:

$$\tau_{\mathbf{q}p}^{\text{film}} = S_{\mathbf{q}p} \tau_{\mathbf{q}p}^0 \tag{A3}$$

For in-plane (parallel) thermal transport, the suppression function is given by [32]:

$$S_{\mathbf{q}p}^{\parallel} = 1 - \frac{(1-\sigma) \text{Kn}_{\mathbf{q}p} \left[ 1 - \exp\left( -\frac{1}{\text{Kn}_{\mathbf{q}p}} \right) \right]}{1 - \sigma \exp\left( -\frac{1}{\text{Kn}_{\mathbf{q}p}} \right)} \tag{A4}$$

Here, $\sigma$ is the specularity parameter, set to zero to model fully diffuse boundary scattering. The Knudsen number $Kn_{\mathbf{q}p}$ is defined as:

$$Kn_{\mathbf{q}p} = \frac{|\Lambda_{\mathbf{q}p}^z|}{L} \tag{A5}$$

where $\Lambda_{\mathbf{q}p}^z$ is the phonon mean free path projected onto the cross-plane (z) direction, and $L$ is the film thickness.

For cross-plane (perpendicular) transport, the suppression function is [32]:

$$S_{\mathbf{q}p}^{\perp} = \frac{1}{1 + 2Kn_{\mathbf{q}p}} \tag{A6}$$

In the resulting thermal conductivity tensor, the $\kappa^{zz}$ component corresponds to the cross-plane thermal conductivity $k_z$, while $\kappa^{xx} = \kappa^{yy}$ represents the in-plane thermal conductivity $k_r$ of the film.

## Acknowledgments

The authors wish to thank Dr. Jun Su from the Center of Optoelectronic Micro & Nano Fabrication and Characterization Facility at the Wuhan National Laboratory for Optoelectronics, Huazhong University of Science and Technology, for his assistance with the SEM and EDX testing. This work is supported by the National Natural Science Foundation of China (NSFC) through Grant No. 52376058.

## Data availability

The data that support the findings of this study are available from the corresponding author upon reasonable request.